\documentclass[10pt,twocolumn,nofootinbib,aps,prd,floatfix]{revtex4-2}
\usepackage{amsmath,amssymb}
\usepackage{mathptmx,mathtools}
\usepackage{bm,booktabs,graphicx,xcolor}
\usepackage[colorlinks=true,linkcolor=blue,citecolor=blue,urlcolor=blue]{hyperref}
\usepackage{microtype}
\usepackage{array}[=2016-10-06]
\usepackage{orcidlink}

\def\nn{\nonumber}
\def\p{\partial}
\def\cN{\mathcal{N}}

\def\cM{\mathcal{M}}

\begin{document}

\title{Multihair thermodynamics of Kerr-Newman-NUT-AdS$_4$ spacetimes}

\author{Di Wu\orcidlink{0000-0002-2509-6729}$^{1,2}$}
\email{Contact author: wdcwnu@163.com, d277wu@uwaterloo.ca}

\author{Shuang-Qing Wu\orcidlink{0000-0001-7936-7195}$^1$}
\email{Contact author: sqwu@cwnu.edu.cn}

\affiliation{$^1$School of Physics and Astronomy, China West Normal University,
Nanchong, Sichuan 637002, People's Republic of China \\
$^2$Department of Physics and Astronomy, University of Waterloo, Waterloo, Ontario N2L 3G1, Canada}

\date{\today}

\begin{abstract}
In this paper, we develop a multihair thermodynamic description of Lorentzian
Kerr-Newman-NUT-AdS$_4$ spacetime in which rotation, electric charge,
AdS pressure and polar conical tensions enter a homogeneous
Christodoulou-Ruffini-type squared-mass formula.  The NUT contribution is
described by a rotation-like secondary hair $J_n=mn/K^2$ and a charge-like
secondary hair $N=n/\sqrt K$.  Both depend on the metric parameters of the
physical solution family but are varied independently in the enlarged
thermodynamic state space.  The mass formula yields the first law and Smarr
relation in a single enthalpy representation, with the area entropy and
Hawking temperature.  The Kerr-Newman-AdS, asymptotically flat
Kerr-Newman-NUT and nonrotating RN-NUT-AdS limits check the angular
normalization and the NUT and pressure contributions.  We also consider
alternative NUT variables, including the conformal dual mass, and derive
the corresponding changes in thermodynamic volume and conical-defect
lengths.  This description applies to the chosen homogeneous state space
and does not require the secondary hairs to be independent asymptotic or
Misner-string surface charges.
\end{abstract}

\maketitle

\section{Introduction}
The choice of thermodynamic variables remains a central issue for
Lorentzian Taub-Newman-Unti-Tamburino (Taub-NUT) spacetimes.  The NUT
parameter, introduced in the Taub solution and the Newman-Unti-Tamburino
extension of Schwarzschild spacetime
\cite{AnnMath53-472,JMP4-915,JMP4-924}, is accompanied by Misner strings,
whose contribution to the first law can be described in several ways.
Rotating AdS solutions also require a choice of angular frame and a
treatment of polar conical deficits and cosmological pressure.  As in
extended AdS black hole thermodynamics, we regard the cosmological
constant as pressure and the mass as enthalpy
\cite{CQG26-195011,CQG28-235017,PRD84-024037,JHEP0712033}.  For NUT
spacetimes, however, a consistent differential first law alone does not
fix how the horizon, string and asymptotic contributions enter thermodynamics.

The multihair formalism introduces thermodynamic secondary hairs to derive
the first law from a Christodoulou-Ruffini-type squared-mass formula
\cite{PRL25-1596,PRD4-3552}.  Applications to four-dimensional Taub-NUT
spacetimes \cite{PRD100-101501} have been extended to dyonic families
\cite{PRD105-124013}.  Comparing the coefficient of the area variation
with the surface gravity gives the usual entropy and temperature
\cite{PRD7-2333,PRD13-191}.
A recent two-horizon diagnostic for uncharged Taub-NUT spacetime
\cite{2607.23644} offers another motivation for the bi-hair description.
Once the sector temperatures are fixed by the inner and outer horizons,
the sum sector closes with the mass and NUT charge, while the difference
sector requires the rotation-like response $J_n=mn$ within the restricted
homogeneous class considered in that work.

Other formulations treat the string contribution explicitly.  Euclidean
and counterterm studies examined the special role of mass and entropy in
NUT/Bolt-AdS geometries
\cite{PRD59-044033,PRD61-084013,hep-th/0305119,1406.1257}, while Lorentzian
treatments use Misner charges, variable string strengths or string
contributions to surface charges
\cite{PRD100-064055,JHEP0719119,CQG36-194001,PLB798-134972,PRD100-104016,
JHEP0520084,JHEP1022044,IJMPD31-2250021,PLB802-135270,PRD106-024022,JHEP0321039}.
Explicit string descriptions are also available for
topological, dyonic and black brane Taub-NUT-AdS systems
\cite{PRD101-124011,PRD108-064022,PRD109-084026,EPJC85-1411}.
Alternatively, string contributions can be absorbed into a modified
thermodynamic mass \cite{PRD105-124034} or a horizon energy
\cite{JHEP1022174,EPJC84-515}.  These formulations address different
macroscopic questions and need not assign the same work terms to the NUT
contribution.

Here we apply the multihair construction to the electrically charged,
rotating AdS family with polar conical deficits.  Previous applications
include RN-NUT-AdS$_4$ solutions in an alternative metric parametrization
\cite{PLB846-138227} and higher even-dimensional Taub-NUT-AdS and RN-NUT-AdS
spacetimes \cite{PRD108-064034,PRD108-064035}.  The present family couples
rotation, electric work, pressure and both polar tensions.  We seek a
squared-mass formula that includes all these contributions while retaining
the area entropy and a homogeneous AdS enthalpy.
There are two geometric difficulties.  First, for nonzero NUT parameter, the
asymptotic angular velocity depends on the polar angle, so the
Kerr-Newman-AdS prescription based on a single nonrotating frame at
infinity \cite{CQG22-1503} cannot be applied directly.  Second, the azimuthal
normalization $K$ controls the polar deficits, whose tensions enter
the first law when the deficits vary \cite{JHEP0517116}.
To account for both effects, we determine the angular frame contribution
together with the mass representation and include the conical tensions
as work variables.

The resulting homogeneous squared-mass formula contains two thermodynamic
secondary hairs associated with the NUT parameter,
\begin{align}
J_n=\frac{mn}{K^2},\quad N=\frac{n}{\sqrt K}.
\end{align}
Their scaling weights are those of angular momentum and electric charge,
respectively, and their conjugates give the rotation-like and charge-like
NUT responses.  Both are functions of the solution parameters on the
physical family.  However, we vary them independently only in the enlarged
thermodynamic state space when differentiating the mass formula.  This
procedure gives the first law and Euler scaling relation without adding
metric parameters or postulating independent asymptotic charges.
The squared-mass formula specifies a homogeneous fundamental relation,
which imposes more structure than differential closure alone.  We check
its restriction to the physical solution family and its Kerr-Newman-AdS,
Kerr-Newman-NUT and RN-NUT-AdS limits.  These checks apply to this mass
representation and do not classify all possible NUT thermodynamics.
Throughout the paper, ``ensemble'' refers to the variables and work terms
of a first-law state space; we do not assume a canonical or grand canonical
boundary potential derived from an on-shell action.
We also examine how this description changes when $N$ is replaced by
$\cN=n/K$ or the conformal dual mass $\tilde M$.  Each choice changes the
variables held fixed in the pressure and tension derivatives.  The volume
and thermodynamic lengths must therefore be specified together with the
NUT variables, even though all these descriptions concern the same
spacetime family.

The remaining part of this paper is organized as follows.
Section \ref{sec:geometry} presents the geometry, polar deficits, horizon
quantities and conformal charges.  Section \ref{sec:multihair} derives the
homogeneous mass formula and its thermodynamic conjugates, then examines
three limiting families.  Section \ref{sec:alternatives} considers
alternative NUT parametrizations and their pressure conjugates.
Section \ref{sec:conclusions} summarizes the results and possible extensions.
Appendix \ref{app:algebra} verifies the mass formula and the differential
identities on the physical solution family.

\section{Kerr-Newman-NUT-AdS$_4$ geometry and thermodynamic quantities}
\label{sec:geometry}

\subsection{Metric and thermodynamic branch}
The Kerr-Newman-NUT-AdS spacetime is a stationary, axially symmetric type D
solution of the Einstein-Maxwell equations with a negative cosmological
constant.  It belongs to the class introduced by Plebanski \cite{AP90-196}
and to the broader Plebanski-Demianski family \cite{AP98-98}.  We choose the
metric gauge in which the Misner strings \cite{JMP4-924} are symmetrically
placed along the two polar axes.  The line element and gauge potential are
\begin{align}
ds^2=&-\frac{\Delta_r}{\Sigma}\left(dt +\frac{2n\cos\theta -a\sin^2\theta}{K}d\phi \right)^2 +\frac{\Sigma dr^2}{\Delta_r}  \nn \\
&+\frac{\Sigma d\theta^2}{\Delta_\theta} +\frac{\Delta_\theta\sin^2\theta}{\Sigma}\left(adt -\frac{r^2 +a^2 +n^2}{K}d\phi\right)^2 \, , \label{KNNUT} \\
A=&\frac{qr}{\Sigma}\left(dt +\frac{2n\cos\theta -a\sin^2\theta}{K}d\phi \right) \, ,
\end{align}
where
\begin{align}
\Delta_r =& \left[1 +g^2\left(r^2 +3n^2\right)\right](r^2+a^2-n^2) \nn \\
&-2mr +q^2 +4g^2n^2r^2 \, , \nn \\
\Delta_\theta =& 1 -g^2(4na\cos\theta +a^2\cos^2\theta) \, ,   \nn \\
\Sigma =& r^2 +(n +a\cos\theta)^2 \, . \nn
\end{align}
Here $m$ and $n$ are the mass and NUT parameters, $a$ is the rotation
parameter, and $q$ is the electric parameter.  The inverse AdS radius is
$g=1/\ell$.  We use units in which Newton's constant is unity and define the
thermodynamic pressure by $P=3g^2/(8\pi)$.  The azimuthal coordinate has period
$2\pi$; its normalization is encoded in $K$, which determines the polar
conical deficits.  We also use $\Xi=1-g^2a^2$.

The standard Kerr-Newman-AdS solution is recovered for $n=0$ and $K=\Xi$,
the asymptotically flat Kerr-Newman-NUT solution for $g=0$ and $K=1$, and
the RN-NUT-AdS family for $a=0$.  The displayed potential contains only the
electric parameter; magnetic charge is not varied in this paper.
We work on a branch with an outer horizon $r_+$ satisfying
$\Delta_r(r_+)=0$, positive horizon area, $K>0$, $\Xi\neq0$ and
$\Delta_\theta>0$ throughout the angular domain.  These conditions define
the local domain used in the thermodynamic calculation; they do not
classify the global causal structure of Lorentzian NUT spacetimes.

\subsection{Polar conical deficits}
Even with symmetrically placed Misner strings, the polar conical tensions
need not be equal: rotation, the NUT parameter and AdS curvature distinguish
the two axes.  We calculate the deficits in local polar patches.  At the
north pole $\theta_+=0$, the shift
$t\rightarrow t-2n\phi/K$ \cite{IJMPD15-335,CQG22-3467} gives
\begin{align}
ds_+^2&=-\frac{\Delta_r}{\Sigma}\left(dt +\frac{2n\cos\theta -2n -a\sin^2\theta}{K}d\phi \right)^2 +\frac{\Sigma dr^2}{\Delta_r} \nn \\
&+\frac{\Sigma d\theta^2}{\Delta_\theta} +\frac{\Delta_\theta\sin^2\theta}{\Sigma}\left(adt -\frac{r^2 +(a +n)^2}{K}d\phi\right)^2 \, ,
\end{align}
On a surface of fixed $t$ and $r$ in this patch, comparison of the leading
azimuthal circumference and proper radial distance near $\theta=0$ gives
the ratio $2\pi\Xi_+/K$, where
\begin{align}
\Xi_+ = \Delta_{\theta=0} = \Xi -4g^2an = 1 -g^2(a^2 +4an) \, .
\end{align}
Similarly, after the shift $t\rightarrow t+2n\phi/K$ and expansion around the
south pole $\theta_-=\pi$, the corresponding ratio is $2\pi\Xi_-/K$, where
\begin{align}
\Xi_- = \Delta_{\theta=\pi} = \Xi +4g^2an = 1 -g^2(a^2 -4an) \, .
\end{align}
The deficits associated with these ratios are \cite{JHEP0517116}
\begin{align}
\delta_\pm = 2\pi\left(1 -\frac{\Xi_\pm}{K} \right) = 2\pi\left[1 -\frac{\Xi \mp 4g^2an}{K}\right]\, ,
\end{align}
and their conical tensions are
\begin{align}
\mu_\pm = \frac{\delta_\pm}{8\pi} = \frac{1}{4}\left[1 -\frac{\Xi \mp 4g^2an}{K}\right] \, . \label{mu}
\end{align}
On the solution family, the tensions depend on $a$, $n$, $g$ and $K$.
Their work terms account for changes in the polar conical geometry, with
variations given by the differential of Eq.~(\ref{mu}).  Including explicit
Misner-string surface charges would require a separate choice of first-law
variables.  The tension difference measures the polar asymmetry,
\begin{align}
\mu_+ -\mu_- &= \frac{2g^2na}{K} \, , \nn
\end{align}
whereas their average measures the common conical deficit,
\begin{align}
\bar{\mu} &= \frac{1}{2}(\mu_+ +\mu_-) = \frac{1}{4}\left(1 -\frac{\Xi}{K} \right) \, . \nn
\end{align}

\subsection{Horizon quantities and angular frame}
We next give the horizon quantities to be compared with derivatives of
the mass formula.  The horizon area and Bekenstein-Hawking entropy are
\begin{align}
\mathcal{A} = 4\frac{\pi(r_+^2 +a^2 +n^2)}{K} \, , \quad S=\frac{\mathcal A}{4}=\frac{\pi(r_+^2+a^2+n^2)}{K} \, . \label{A}
\end{align}
The surface gravity is
\begin{align}
\kappa = \frac{\Delta_r^{'}|_{r=r_+}}{2(r_+^2 +a^2 +n^2)} \, , \label{kappa}
\end{align}
where the event horizon is determined by $\Delta_r|_{r=r_+}=0$.
The horizon generator is
\begin{align}
\chi = \p_t +\Omega_H\p_\phi \, ,
\end{align}
where the horizon angular velocity is
\begin{align}
\Omega_H = -\frac{g_{t\phi}}{g_{\phi\phi}}\Big|_{r = r_+} = \frac{aK}{r_+^2 +a^2 +n^2} \, . \label{Omegah}
\end{align}
The asymptotic electric charge follows from the Gauss-law integral,
\begin{align}
Q = \frac{1}{4\pi}\int\star F = \frac{q}{K} \, , \label{Q}
\end{align}
and the electrostatic potential difference is
\begin{align}
\Phi = A_\mu\chi^\mu\big|_{r = r_+} -A_\mu\chi^\mu\big|_{r\rightarrow \infty} = \frac{qr_+}{r_+^2 +a^2 +n^2} \, . \label{Phi}
\end{align}
The charge integral is evaluated at infinity, and the potential difference
is taken along the horizon generator in the chosen gauge.  Thus the
electric normalization is fixed independently of the thermodynamic angular
frame: $\Phi$ involves $\Omega_H$, while the angular work term involves
the enthalpy conjugate $\Omega$.

For Kerr-Newman-AdS black holes, the angular velocity entering the first law
is measured relative to a nonrotating frame at infinity \cite{CQG22-1503},
\begin{align}
\Omega = \Omega_H -\Omega_\infty \, ,
\end{align}
where
\begin{align}
\Omega_\infty &= -\frac{g_{t\phi}}{g_{\phi\phi}}\Big|_{r\rightarrow\infty} \nn \\
&= \frac{g^2(2n\cos\theta -a\sin^2\theta)K}{\left[-g^2\left(2n\cos\theta -a\sin^2\theta \right)^2 +\Delta_\theta\sin^2\theta \right]} \, .
\end{align}
For nonzero NUT parameter, however, this expression depends on $\theta$, so a
constant angular shift cannot remove the boundary rotation at every angle.
We therefore derive the thermodynamic angular velocity from the homogeneous
enthalpy in section \ref{sec:multihair}.  Its frame term ensures that the
Kerr-Newman-AdS normalization is recovered as $n\to0$.

\subsection{Conformal charges and thermodynamic mass}
To distinguish this thermodynamic choice from the asymptotic charges,
we first evaluate the conformal mass $\cM$ and angular momentum $J$ using
the Ashtekar-Magnon-Das method \cite{PRD73-104036,PRD110-104021}.
The leading metric after conformal rescaling is
\begin{align}
ds_\infty^2 =& -g^2\left[dt +\frac{(2n\cos\theta -a\sin^2\theta)d\phi}{K} \right]^2 \nn \\
&+\frac{dr^2}{g^2r^4} +\frac{d\theta^2}{\Delta_\theta} +\frac{\Delta_\theta\sin^2\theta d\phi^2}{K^2} \, .
\end{align}
The conserved charge $Q[\xi]$ associated with a Killing vector $\xi$ is
\begin{align}
&Q[\xi] = \frac{1}{8\pi g^3}\int rN^\alpha N^\beta C^{\mu}_{\alpha\nu\beta}\xi^{\nu}dS_\mu \, ,
\end{align}
where
\begin{align}
dS_\mu = \frac{g\sin\theta}{K}d\theta \wedge d\phi
\end{align}
is the surface element on the two-sphere section of the conformal boundary.
The resulting mass and angular momentum are
\begin{align}
&\cM = Q[\xi^{t}] = \frac{m}{K} \, , \\
&J = Q[\xi^{\phi}] = \frac{ma}{K^2} \, . \label{J}
\end{align}
The conformal dual charges will be useful for the alternative NUT
parametrization in section \ref{sec:alternatives}.  With the dual Weyl tensor
\begin{align}
\tilde{C}_{\mu\nu\rho\sigma} = \frac{1}{2}\epsilon_{\mu\nu\alpha\beta}C^{\alpha\beta}_{~~~\rho\sigma} \, ,
\end{align}
where $\epsilon_{\mu\nu\rho\sigma}$ is the Levi-Civita tensor, one obtains
\begin{align}
&\tilde{M} = \tilde{Q}[\xi^{t}] = \frac{n}{K}\left[1 -g^2(a^2 -4n^2)\right] \, , \label{tM} \\
&\tilde{J} = \tilde{Q}[\xi^{\phi}] = \frac{na}{K^2}\left[1 -g^2(a^2 -4n^2)\right] \, .
\end{align}
The electric and magnetic AMD charges of the dyonic Kerr-Newman-NUT-AdS
family, including their dependence on the Misner-string position, were
studied in Ref. \cite{PRD110-104021}.  The expressions above use the symmetric
string placement and the azimuthal normalization of the present metric.

The homogeneous mass formula below uses $M=m/(\Xi K)$ as the enthalpy.
It differs from $\cM=m/K$ by a term proportional to $J$ arising from the
thermodynamic angular frame.  The mass and angular potential entering the
first law must refer to the same representation.  As shown in Appendix
\ref{app:algebra}, for a generic rotating AdS solution, substituting $\cM$ into this
squared-mass formula does not give equality between its two sides.  The formula
therefore fixes its own mass normalization without changing the AMD charge
assignment.

\section{Multihair thermodynamics}
\label{sec:multihair}
\subsection{Thermodynamic state space and NUT secondary hairs}\label{CFL}

We construct the mass formula in an enlarged thermodynamic state space
and then restrict it to the physical solution family.  The physical
parameters $(m,a,n,q,g,K,r_+)$ satisfy the horizon equation and admit six
independent local variations.  In differentiating the fundamental relation,
we treat $(S,J,J_n,Q,N,P,\mu_+,\mu_-)$ as independent variables.  We then
express their differentials in terms of variations of the physical
parameters.  This restriction is the pullback to the solution family.

The homogeneous mass formula includes the horizon, rotation, electric,
NUT, pressure and polar-deficit contributions with their natural scaling
weights and the limits discussed below.  It also specifies the extension
away from the physical family: we add no terms that vanish on that family
but change the derivatives away from it.  The thermodynamic derivatives
are thus fixed by the chosen formula; homogeneity alone would not determine
the extension.

The rotation-like NUT secondary hair is
\begin{align}
J_n = \frac{mn}{K^2} \, , \label{J_n}
\end{align}
and the charge-like NUT variable is
\begin{align}
N = \frac{n}{\sqrt{K}} \, , \label{Q_n}
\end{align}
Their scaling weights are two and one, respectively.  The factor
$\sqrt K$ in $N$ simplifies the conical normalization in the mass formula.
Both quantities depend on the metric parameters of the physical family,
so treating them as independent thermodynamic variables introduces neither
new solutions nor independent asymptotic charges.  Including the angular
frame contribution gives the enthalpy
\begin{align}
M = \cM +\frac{g^2a}{\Xi}KJ = \frac{m}{\Xi K} \, . \label{M}
\end{align}

\subsection{Homogeneous squared-mass formula}
The fundamental relation is
$M=M(S,J,J_n,Q,N,P,\mu_+,\mu_-)$.
For compactness we express it in terms of $\mathcal A=4S$.
Under a rescaling of length, the variables have weights
$[S]=[\mathcal A]=[J]=[J_n]=2$, $[Q]=[N]=1$, $[P]=-2$ and
$[\mu_\pm]=0$, while $M$ has weight one.  As for the Kerr-Newman and
Kerr-Newman-AdS mass formulae \cite{PRL25-1596,PRD4-3552,CQG17-399},
one homogeneous relation then generates both the differential first law
and the Smarr scaling relation \cite{CMP31-161,PRL30-71}.

To construct this relation, we eliminate the mass parameter using
$\Delta_r(r_+)=0$ and rewrite the horizon data in thermodynamic variables.
Define $B=1-2\mu_+-2\mu_-$.  On the physical family,
$B=\Xi/K$ and $\mathcal A/(4\pi)-2N^2=(r_+^2+a^2-n^2)/K$.
Together with $\mu_+-\mu_-=2g^2an/K$, these identities reduce the
expression $X$ below to $2mr_+/K^2$.  The resulting squared-mass formula is
\begin{align}
B\frac{M^2\mathcal A}{4\pi}
=\frac{X^2}{4}+J^2\left(1+\frac{2P\mathcal A}{3B}\right)+J_n^2\, ,\label{smf}
\end{align}
where the full expression for $X$ is
\begin{align}
&X=\left(B+\frac{32}{3}\pi PN^2\right)
\left(\frac{\mathcal A}{4\pi}-2N^2\right)+Q^2\nn\\
&\quad+\frac{P\mathcal A^2}{6\pi}
-\frac{3(\mu_+-\mu_-)^2}{8\pi P}\, .\label{Xdef}
\end{align}
Appendix \ref{app:algebra} verifies Eq.~(\ref{smf}) on the physical family
using the horizon equation.  Treating it as a fundamental relation also
defines its extension away from that family.  The last term in $X$
restricts this extension to $P>0$.  We take the flat limit only after
restricting to the physical family: $P=3g^2/(8\pi)$ and
$\mu_+-\mu_-=2g^2an/K$ imply that $(\mu_+-\mu_-)^2/P$ vanishes as $g\to0$.

\subsection{Thermodynamic conjugates and scaling relation}
\label{sec:conjugates}
Differentiating Eq.~(\ref{smf}), with $S=\mathcal A/4$, gives the first law
\begin{align}
dM =& TdS +\Omega dJ +\omega{}dJ_n +\Phi dQ +\Psi dN +VdP \nn \\
& -\lambda_+d\mu_+ -\lambda_-d\mu_- \, . \label{FL}
\end{align}
Each coefficient is a partial derivative in the enlarged state space,
evaluated on the solution family.  All derivatives below hold the remaining
variables in $(S,J,J_n,Q,N,P,\mu_+,\mu_-)$ fixed, with $S=\mathcal A/4$.
In particular, the entropy derivative gives the geometric Hawking temperature,
\begin{align}
&T = \frac{\p M}{\p S} = \frac{\kappa}{2\pi}
= \frac{r_+\left[1 +g^2\left(2r_+^2 +a^2 +6n^2\right)\right] -m}{2\pi(r_+^2 +a^2 +n^2)} \, . \label{ts}\end{align}
The angular velocity obtained from the squared-mass formula is
\begin{align}
&\Omega = \frac{\p M}{\p J} = \frac{4\pi J}{M\mathcal{A}B} +\frac{8\pi PJ}{3MB^2} = \frac{aK}{r_+^2 +a^2 +n^2} +\frac{g^2aK}{\Xi}\, , \label{Omega}\end{align}
where holding $\mathcal A$ fixed is equivalent to holding $S$ fixed.
The first term is the horizon angular velocity in Eq.~(\ref{Omegah}).
The second is the frame contribution conjugate to the angular momentum
in the chosen enthalpy representation.  By continuity with the $n=0$
convention, we denote the corresponding constant subtraction by
\begin{align}
\Omega_\infty = -\frac{g^2aK}{\Xi} \, .
\end{align}
For $n\neq0$, however, this constant is distinct from the angle-dependent asymptotic
quantity displayed in section \ref{sec:geometry}.  The potential conjugate
to $J_n$ also has the dimensions of angular velocity:
\begin{align}
&\omega{} = \frac{\p M}{\p J_n} = \frac{4\pi J_n}{M\mathcal{A}B}
= \frac{nK}{r_+^2 +a^2 +n^2} \, . \label{Omega_n}\end{align}
The electric potential is
\begin{align}
&\Phi = \frac{\p M}{\p Q} = \frac{2\pi QX}{M\mathcal{A}B} = \frac{qr_+}{r_+^2 +a^2 +n^2} \, . \end{align}
The NUT potential conjugate to $N$ is
\begin{align}
\Psi &= \frac{\p M}{\p N} = -\frac{4\pi NX}{M\mathcal{A}} +\frac{16\pi PNX(\mathcal{A} -16\pi N^2)}{3M\mathcal{A}B} \nn \label{Psi} \\
&= -\frac{2nr_+\left[1 -g^2\left(2r_+^2 +3a^2 -6n^2 \right)\right]}{\sqrt{K}(r_+^2 +a^2 +n^2)} \, .
\end{align}
For the pressure response it is convenient to introduce
\[
\mathcal{Y}_P =
\frac{\mathcal{A}^2}{32\pi} +\frac{\mathcal{A}N^2}{2} -4\pi N^4
+\frac{9\left(\mu_+ -\mu_- \right)^2}{128\pi P^2} \, .
\]
in terms of which the volume is written as
\begin{align}
V &= \frac{\p M}{\p P} = \frac{16\pi X\mathcal{Y}_P}{3M\mathcal{A}B}
+\frac{4\pi J^2}{3MB^2} \nn \label{tV} \\
&= \frac{4\pi}{3K}\left\{\frac{ma^2}{\Xi} +r_+\left[r_+^2 +a^2 +5n^2 +\frac{4n^2(a^2 -2n^2)}{r_+^2 +a^2 +n^2} \right] \right\} \, .
\end{align}
\begin{samepage}
The thermodynamic lengths are conjugate to the polar tensions, with signs
set by the convention $-\lambda_\pm d\mu_\pm$ in Eq.~(\ref{FL}):
\begin{align}
\lambda_\pm =& -\frac{\p M}{\p \mu_\pm} = \frac{1}{MB}\Bigg\{-\frac{8\pi PJ^2}{3B^2} -M^2 \nn \label{lambda} \\
&+\frac{X}{2}\Bigg[1 -\frac{8\pi N^2}{\mathcal{A}} \pm\frac{3(\mu_+ -\mu_-)}{2P\mathcal{A}} \Bigg] \Bigg\} \nn \\
=&~ r_+ -\frac{m(2 -\Xi)}{\Xi^2} \pm\frac{2n(a \mp n)r_+}{r_+^2 +a^2 +n^2}  \, .
\end{align}
\end{samepage}
The volume in Eq.~(\ref{tV}) is defined at fixed $N$ and fixed values of
the other specified variables.  Changing the NUT variable therefore changes
the pressure derivative, as discussed in section \ref{sec:alternatives}.

Euler scaling gives the Bekenstein-Smarr relation
\begin{align}
M = 2TS +2\Omega J +2\omega{}J_n +\Phi Q +\Psi N -2VP \, , \label{Smarr}
\end{align}
The tensions do not appear explicitly because their scaling weight is zero,
although their variations contribute to the first law.
On the physical family, Eqs. (\ref{FL}) and (\ref{Smarr}) follow upon using
the horizon equation and its differential.  Appendix \ref{app:algebra}
gives these identities in terms of the physical parameters, before any
limiting family is selected.

This mass representation is therefore compatible with $S=\mathcal A/4$
and $T=\kappa/(2\pi)$.  The NUT work is expressed through $J_n$ and $N$,
and the conical work through $\mu_\pm$ and their conjugate lengths.
Table \ref{tab:nut-ensembles} compares this description with formulations
using explicit Misner/string variables
\cite{PRD100-064055,JHEP0719119,CQG36-194001,PLB798-134972,IJMPD31-2250021}
or a different energy assignment.
\begin{table*}[t]
\centering
\caption{Macroscopic organizations of NUT thermodynamics.  The rows specify the energy and work variables used to formulate each thermodynamic problem.}
\label{tab:nut-ensembles}
\small
\setlength{\tabcolsep}{3pt}
\begin{tabular}{
>{\raggedright\arraybackslash}p{0.14\textwidth}
>{\raggedright\arraybackslash}p{0.21\textwidth}
>{\raggedright\arraybackslash}p{0.30\textwidth}
>{\raggedright\arraybackslash}p{0.25\textwidth}
}
\hline
Formulation & Energy variable & NUT/string data & Question addressed \\
\hline
Multihair & $M=m/(\Xi K)$ in the homogeneous enthalpy representation &
$S=\mathcal A/4$, $J_n$, $N$ and the conical tensions $\mu_\pm$ &
How do the response terms enter a homogeneous squared-mass formula? \\
\hline
Misner/string & Energy associated with the chosen surface-charge prescription &
Explicit string charges or variable string strengths &
How does the string sector contribute to the first law? \\
\hline
Mass choices & Modified spacetime mass or horizon energy &
String contributions absorbed into the energy or its conjugates &
How is the energy assigned to the spacetime or horizon? \\
\hline
\end{tabular}
\end{table*}
\subsection{Limiting families}
We check the angular normalization, the flat NUT limit and the NUT
contribution to the pressure response in turn.  In each case, we first
differentiate the full mass formula and then restrict the conjugates to
the limiting family.  Comparison with the standard solutions also requires
the conical normalizations specified below.

\paragraph{Kerr-Newman-AdS limit.}
For $n=0$ and $K=\Xi$, the NUT parameter and both conical deficits vanish.
The surviving quantities are
\begin{align}
\begin{aligned}
&S = \frac{\pi(r_+^2 +a^2)}{\Xi} \, ,\quad \Phi = \frac{qr_+}{r_+^2 +a^2} \, , \\
&M = \frac{m}{\Xi^2} \, , \quad J = \frac{ma}{\Xi^2} \, , \quad Q = \frac{q}{\Xi} \, ,\quad \Omega = \frac{a\Xi}{r_+^2 +a^2} +g^2a \, ,  \\
&V = \frac{4\pi}{3\Xi}\left[\frac{ma^2}{\Xi} +r_+(r_+^2 +a^2)\right] \, ,\quad K = \Xi = \Xi_\pm = 1 -g^2a^2 \, ,  \\
&N = \Psi = \tilde{M} = J_n = \omega{} = \mu_\pm = 0 \, . \end{aligned}
\label{n=0}
\end{align}
These are the standard Kerr-Newman-AdS quantities in the nonrotating
thermodynamic frame.  The condition $K=\Xi$ selects the solution without
conical deficits; it is an additional normalization, not a consequence of
$n=0$ alone.  Both NUT secondary hairs and their work terms vanish in this
limit.

\paragraph{Asymptotically flat Kerr-Newman-NUT limit.}
Taking $g\to0$ on the physical family and setting $K=1$ gives
\begin{align}
\begin{aligned}
&P = \mu_+ = \mu_- = 0 \, , \quad \Xi_+ = \Xi_- = \Xi = K = 1 \, ,  \\
&\tilde{M} = N = n \, ,\quad M = m \, , \quad Q = q \, , \quad J = ma \, , \quad  J_n = mn \, ,  \\
&\Psi = -\frac{2nr_+}{r_+^2 +a^2 +n^2} \, ,\quad S = \pi(r_+^2 +a^2 +n^2) \, ,  \\
&\Omega = \frac{a}{r_+^2 +a^2 +n^2} \, , \\
&\Phi = \frac{qr_+}{r_+^2 +a^2 +n^2} \, , \quad \omega{} = \frac{n}{r_+^2 +a^2 +n^2} \, .
\end{aligned}
\label{g=0}
\end{align}
The resulting expressions agree with the flat Kerr-Newman-NUT multihair
construction of Ref. \cite{PRD100-101501}.  Pressure and conical-deficit
variations are absent within this limiting family, while $J_n=mn$ and
$N=n$ retain the rotation-like and charge-like NUT responses.

\paragraph{Nonrotating RN-NUT-AdS limit.}
For $a=0$ and $K=1$, the angular frame contribution and conical deficits
vanish, while the charged NUT and pressure sectors remain:
\begin{align}
\begin{aligned}
&\mu_+ = \mu_- = \Omega = J = 0 \, ,\quad \Xi_+ = \Xi_- = \Xi = K = 1 \, ,  \\
&M = m \, , \quad Q = q \, , \quad J_n = mn \, , \quad N = n \, ,  \\
&\tilde{M} = n(1 +4g^2n^2) \, ,\quad S = \pi(r_+^2 +n^2) \, , \quad \Phi = \frac{qr_+}{r_+^2 +n^2} \, ,  \\
&\omega{} = \frac{n}{r_+^2 +n^2} \, ,\quad \Psi = -\frac{2nr_+}{r_+^2 +n^2}\left[1 -2g^2(r_+^2 -3n^2)\right] \, ,  \\
&V = \frac{4}{3}\pi r_+\left(r_+^2 +5n^2 -\frac{8n^4}{r_+^2 +n^2}\right) \, .
\end{aligned}
\label{a=0}
\end{align}
In this family, the NUT contribution to the volume can be examined without
the rotating frame term.  The volume is conjugate to pressure in the primary
$(\Psi,N)$ representation and differs from those obtained with other NUT
variables.  We examine this dependence in the next section.  The three
limits above check the angular, NUT and pressure normalizations of the full
mass formula.

\section{Alternative parametrizations of the NUT sector}
\label{sec:alternatives}
We now change the variable used for the charge-like NUT response while
retaining $J_n$.  We compare $N$ with $\cN=n/K$ and the conformal dual
mass $\tilde M$.  Their different dependence on $K$, $a$ and $g$ changes
the pressure and conical-tension coefficients as well as the NUT potential.

Applying the chain rule on the physical solution family preserves the
total differential identity and gives the corresponding Smarr relation.
This reparametrization, however, need not define an independent extension of the
compact squared-mass formula to unconstrained variables.
Table \ref{tab:nut-parametrizations} lists the three choices.

\begin{table*}[t]
\centering
\caption{Charge-like NUT variables and their pressure conjugates.
The rotation-like secondary hair $J_n$ is retained in all three descriptions.}
\label{tab:nut-parametrizations}
\small
\setlength{\tabcolsep}{9pt}
\renewcommand{\arraystretch}{1.18}
\begin{tabular}{llll}
\hline
NUT variable & Conjugate & Pressure conjugate & Role in this paper \\
\hline
$N=n/\sqrt K$ & $\Psi$ & $V$ & homogeneous mass representation \\
$\mathcal N=n/K$ & $\bar\Psi$ & $\bar V$ & alternative NUT normalization \\
$\tilde M$ & $\tilde\Psi$ & $\tilde V$ & conformal dual-mass variable \\
\hline
\end{tabular}
\end{table*}

\subsection{The \texorpdfstring{$\mathcal N=n/K$}{N=n/K} parametrization}
The first alternative uses
\begin{align}
\cN = \frac{n}{K} \label{bQn}
\end{align}
in place of $N=n/\sqrt K$.  Its azimuthal normalization coincides with that
of the electric charge and the overall normalization of the dual mass.
The first law and Smarr relation take the form
\begin{align}
dM =& TdS +\Omega dJ +\omega{}dJ_n +\Phi dQ +\bar{\Psi}d\cN +\bar{V}dP \nn \\
& -\bar{\lambda}_+d\mu_+ -\bar{\lambda}_-d\mu_- \, , \\
M =& 2TS +2\Omega J +2\omega{}J_n +\Phi Q +\bar{\Psi}\cN -2\bar{V}P \, . \end{align}
Expressing $dN$ in terms of $d\cN$, $dP$ and $d\mu_\pm$ on the physical
family gives the shifted coefficients
\begin{align}
\begin{aligned}
&\bar{\Psi} = \frac{\sqrt{K}}{\Xi}\Psi \, , \quad \bar{V} = V +\frac{4\pi}{3K}a^2n\bar{\Psi} \, , \\
&\bar{\lambda}_\pm = \lambda_\pm \pm \left(\frac{a}{2} \mp n\right)\bar{\Psi} \, , \end{aligned}
\label{LVP1}
\end{align}
where the unbarred quantities are defined in section \ref{sec:conjugates}.
The shifts in $\bar V$ and $\bar\lambda_\pm$ account for the change of
NUT variable.  They preserve the total work while redistributing it among
the NUT, pressure and conical terms.

\subsection{The dual mass parametrization}
A second choice is the conformal dual mass in Eq.~(\ref{tM}).  Unlike $N$
or $\cN$, it contains an explicit curvature-dependent factor.  Denoting its
conjugate by $\tilde\Psi$, we obtain
\begin{align}
dM =& TdS +\Omega dJ +\omega{}dJ_n +\Phi dQ +\tilde{\Psi} d\tilde{M} +\tilde{V}dP \nn \\
&-\tilde{\lambda}_+d\mu_+ -\tilde{\lambda}_-d\mu_- \, , \\
M =& 2TS +2\Omega J +2\omega{}J_n +\Phi Q +\tilde{\Psi}\tilde{M} -2\tilde{V}P \, , \end{align}
with
\begin{align}
\begin{aligned}
&\tilde{V} = V -\frac{4\pi n}{3K}\left\{8n^2 +a^2\left[1 +g^2(4n^2 -a^2) \right]\right\}\tilde{\Psi} \, , \\
&\tilde{\lambda}_\pm = \lambda_\pm \mp \frac{1}{2}(a \pm 2n)\left[1 -g^2\left(a\mp 2n \right)^2 \right]\tilde{\Psi} \, , \\
&\tilde{\Psi} = \frac{\sqrt{K}}{1 +12g^2n^2 +g^4a^2(4n^2 -a^2)}\Psi \, ,
\end{aligned}
\label{LVP2}
\end{align}
where the unshifted quantities are defined in section \ref{sec:conjugates}.
This parametrization applies when the denominator of $\tilde\Psi$ is
nonzero.  Its pressure-dependent normalization produces the shift in
$\tilde V$, and its dependence on the polar geometry changes the lengths.
The volumes $V$, $\bar V$ and $\tilde V$ thus correspond to different
conditions on the NUT variable.  We next consider a mixed description
that allows comparison with a specified nonrotating volume.

\subsection{Mixed NUT variables and thermodynamic volume}
Comparing volumes requires specifying the NUT work terms.  We can
redistribute these terms more generally by using $N$, $\cN$ and $\tilde M$
together, subject to their relations on the physical family.  This does
not introduce three additional independent solution parameters.

We use this freedom to examine the Lorentzian continuation of the
Taub-Bolt volume in Ref. \cite{CQG31-235003}.  That calculation uses
Euclidean NUT conventions.  Accordingly, our comparison is with the analytically
continued volume and does not identify the Euclidean and Lorentzian
ensembles.  We denote the redistributed NUT potentials by lowercase
symbols to distinguish them from $\Psi$, and write
\begin{widetext}
\begin{align}
-\Psi N +2VP &= -(\psi N +\bar{\psi}\cN +\tilde{\psi}\tilde{M}) + 2\mathcal{V}P \, , \nn \\
\Psi dN +VdP -\lambda_+d\mu_+ -\lambda_-d\mu_- &= \psi dN +\bar{\psi}d\cN +\tilde{\psi}d\tilde{M} +\mathcal{V}dP -L_+d\mu_+ -L_-d\mu_- \, . \nn
\end{align}
We split the physical expressions for $\Psi$ and $V$ in
Eqs. (\ref{Psi}) and (\ref{tV}) as
\begin{align}
&\Psi = \frac{-2nr_+[1 +3(2w-1)g^2a^2]}{\sqrt{K}(r_+^2 +a^2 +n^2)} +\frac{4g^2nr_+(r_+^2 +3wa^2 -3n^2)}{\sqrt{K}(r_+^2 +a^2 +n^2)}  \, , \label{Pn}\\
&V = \frac{4\pi ma^2}{3\Xi K} +\frac{4\pi r_+}{3K}\left[r_+^2 +a^2 +3n^2 +\frac{6(1-w)a^2n^2}{r_+^2 +a^2 +n^2} \right] +\frac{8\pi n^2r_+}{3K}Y \, , \label{V}
\end{align}
where
\begin{align}
Y &= \frac{r_+^2 +3wa^2 -3n^2}{r_+^2 +a^2 +n^2} \, , \nn
\end{align}
and $w$ is a constant, with $w=0$ and $w=1$ providing two convenient
choices.  The corresponding mixed coefficients are
\begin{align}
\begin{aligned}
&\psi  = -\frac{2nr_+\left[1 +3(2w-1)g^2a^2 \right]}{\sqrt{K}(r_+^2 +a^2 +n^2)} \, ,\quad \tilde{\psi} = \frac{2nr_+}{8n^2 +a^2(\Xi +4g^2n^2)}Y \, ,  \\
&\bar{\psi} = -\frac{1 +12g^2n^2 +g^4a^2(4n^2 -a^2)}{\Xi}\tilde{\psi} +\frac{\sqrt{K}(\Psi -\psi )}{\Xi} \, ,\quad L_\pm = \lambda_\pm +\frac{8g^2n^2(2n \mp a)\mp a\Xi}{\Xi}\tilde{\psi} +\frac{\pm a -2n}{2\Xi}(\Psi -\psi)\sqrt{K} \, ,  \\
&\mathcal{V} = V - \frac{8\pi}{3\Xi K}\left[4n^2 +a^2(\Xi +4g^2n^2)\right]n\tilde{\psi} +\frac{4\pi na^2}{3\Xi\sqrt{K}}(\Psi -\psi) \, .
\end{aligned}
\label{LVP3}
\end{align}
\end{widetext}
The parameter $w$ specifies the redistribution of NUT work.  It does not
enter the metric or change the number of independent physical variations.
For $a=0$ and $K=1$, the volume becomes
\begin{align}
\mathcal{V} = \frac{4\pi r_+}{3}(r_+^2 +3n^2) \, ,
\end{align}
which has the Lorentzian form obtained by continuing the Euclidean
Taub-Bolt volume of Ref. \cite{CQG31-235003}.
A two-pair description with $\bar\psi=0$ can reproduce the same
nonrotating volume.  To obtain this description, we retain $\tilde\psi$
from Eq.~(\ref{LVP3}) and
reassign $\psi$ so that the coefficient of $d\cN$ vanishes.
The NUT work is then organized as
\begin{widetext}
\begin{align}
&-\Psi N +2VP = -(\psi N +\tilde{\psi}\tilde{M}) + 2\mathcal{V}P \, ,\quad \Psi dN +VdP -\lambda_+d\mu_+ -\lambda_-d\mu_- = \psi dN +\tilde{\psi}d\tilde{M} +\mathcal{V}dP -L_+d\mu_+ -L_-d\mu_- \, . \nn \end{align}
The required shifts in the NUT potential, lengths and volume are
fixed together by the differential of $\tilde M$, giving
\begin{align}
\begin{aligned}
&\psi = \Psi -\frac{1 +12g^2n^2 +g^4a^2(4n^2 -a^2)}{\sqrt{K}}\tilde{\psi} \, ,\quad \tilde{\psi} = \frac{2nr_+}{8n^2 +a^2(\Xi +4g^2n^2)}Y \, ,\quad L_\pm = \lambda_\pm -\frac{1}{2}(\pm a +2n)\left[1 -g^2(a \mp 2n)^2 \right]\tilde{\psi} \, ,  \\
&\mathcal{V} = V -\frac{4\pi n}{3K}\left[8n^2 +a^2(\Xi +4g^2n^2) \right]\tilde{\psi} = \frac{4\pi ma^2}{3\Xi K}
+\frac{4\pi r_+}{3K}\left[r_+^2 +a^2 +3n^2
+\frac{6(1-w)a^2n^2}{r_+^2 +a^2 +n^2}\right] \, . \end{aligned}
\label{LVPtwo}
\end{align}
\end{widetext}
These expressions apply where $8n^2+a^2(\Xi+4g^2n^2)\neq0$.
They satisfy both displayed work identities on the physical family.
For $a=0$ and $K=1$, the volume reduces to
\begin{align}
\mathcal{V} = \frac{4\pi r_+}{3}(r_+^2 +3n^2) \, ,
\end{align}
as in the three-pair case.  Equation~(\ref{LVPtwo}) therefore gives a
two-pair representation of this volume that preserves the first law and
Smarr relation.  It follows from keeping $\tilde\psi$ and imposing
$\bar\psi=0$; a different redistribution need not give the same pressure
conjugate.  A thermodynamic volume must consequently be specified together
with the NUT and conical work terms.  Agreement or disagreement between
the pressure coefficients alone is insufficient to establish equivalence
or inconsistency of the full thermodynamic descriptions.

\section{Conclusions}
\label{sec:conclusions}
We have derived a homogeneous Christodoulou-Ruffini-type squared-mass
formula (\ref{smf}) for Lorentzian Kerr-Newman-NUT-AdS$_4$ spacetime with
symmetric Misner strings and polar conical deficits.  The formula includes
rotation, electric charge, AdS pressure and both conical tensions in a
single enthalpy representation.  The NUT contribution enters through the
rotation-like secondary hair $J_n=mn/K^2$ and the charge-like variable
$N=n/\sqrt K$, both of which have nonzero work terms in the first law.
Their dependence on the metric parameters is restored when the enlarged
thermodynamic state space is restricted to the physical family.

The mass formula gives both the differential first law and the Euler
scaling relation, with the area entropy and Hawking temperature.  Its
Kerr-Newman-AdS, flat Kerr-Newman-NUT and nonrotating RN-NUT-AdS limits
check the angular normalization, NUT response and pressure contribution,
respectively.  The enthalpy $M=m/(\Xi K)$ and its angular conjugate are
those of this mass representation, while the AMD conformal mass remains
an asymptotic charge.  This construction gives a specific homogeneous
description of NUT thermodynamics; it does not establish uniqueness among
all choices of energy and string variables.

Replacing the NUT variable by $\cN=n/K$ or $\tilde M$ redistributes the
pressure and conical-defect work according to the chain rule.  The volume
and thermodynamic lengths must therefore be given together with the
variables held fixed in their definition.  We have also constructed a
mixed two-pair description that reproduces the chosen nonrotating volume
and preserves the first law and Smarr relation.  Thus, comparisons between
thermodynamic formulations of a NUT spacetime must account for the full
first law, not just individual work coefficients.

Including magnetic charge would allow us to study how the dyonic sector
enters the squared-mass formula.  The resulting state space could also be
used to study phase structure and thermodynamic topology, following the
NUT applications \cite{EPJC83-365,EPJC83-589} and the broader classification
programme \cite{PRL129-191101,PRD107-024024,PRD107-084002,PRD108-084041,
JHEP0624213,PLB856-138919,PDU46-101617,CQG42-125007,PLB860-139163,PRD110-L081501,
PRD111-L061501,PLB865-139482,EPJC85-828,PRD112-124024,EPJC86-187,2602.05231,2605.00037}.
These applications require a separate choice of ensemble and stability
analysis beyond the first-law calculation given here.

\begin{acknowledgments}
This work is supported by the National Natural Science Foundation of China
(NSFC) under Grants No. 12675077, No. 12205243 and No. 12375053, the China Scholarship
Council (CSC), and the Sichuan Science and Technology Program under Grant
No. 2026NSFSC0021.

\end{acknowledgments}

\appendix

\section{Algebraic identities and physical variations}
\label{app:algebra}

We verify the mass formula and its differential on the physical
Kerr-Newman-NUT-AdS$_4$ family.  We first differentiate the squared-mass
formula with independent thermodynamic variables, then use the horizon
equation and the definitions of the physical quantities to evaluate the
resulting expressions on the solution family.

\subsection{Squared-mass formula from the horizon equation}

The horizon equation $\Delta_r(r_+)=0$ can be written as
\begin{align}
2mr_+ = \left[1+g^2(r_+^2+3n^2)\right](r_+^2+a^2-n^2) +q^2+4g^2n^2r_+^2 \, . \label{app:horizon}
\end{align}
Using
\begin{align}
&B=\frac{\Xi}{K}\,,\quad \frac{\mathcal A}{4\pi}-2N^2=\frac{r_+^2+a^2-n^2}{K}\,, \nn  \\
&
\mu_+-\mu_-=\frac{2g^2an}{K}\,,
\end{align}
the bracket $X$ in Eq.~(\ref{Xdef}) becomes
\begin{align}
X=\frac{2mr_+}{K^2}
\end{align}
on the physical branch.  The left side of Eq.~(\ref{smf}) is
\begin{align}
B\frac{M^2\mathcal A}{4\pi}
=\frac{m^2(r_+^2+a^2+n^2)}{\Xi K^4}\,,
\end{align}
while the right side gives
\begin{align}
\frac{X^2}{4}
+J^2\left(1+\frac{2P\mathcal A}{3B}\right)+J_n^2 =& \frac{m^2a^2}{K^4}\left[1+\frac{g^2(r_+^2+a^2+n^2)}{\Xi}\right] \nn  \\
&+\frac{m^2r_+^2}{K^4} +\frac{m^2n^2}{K^4}\nn \\
 =& \frac{m^2(r_+^2+a^2+n^2)}{\Xi K^4}\,,
\end{align}
where $\Xi=1-g^2a^2$ was used in the last step.  This verifies the
restriction of the squared-mass formula to the horizon equation.
For comparison, replacing $M$ by the AMD conformal mass
$\mathcal M=m/K$ in the same expression gives
\begin{align}
B\frac{\mathcal M^2\mathcal A}{4\pi}
=\frac{\Xi m^2(r_+^2+a^2+n^2)}{K^4}\,.
\end{align}
Subtracting this expression from the right-hand side of Eq.~(\ref{smf}) gives
\begin{align}
\frac{m^2(r_+^2+a^2+n^2)}{K^4}\left(\frac{1}{\Xi}-\Xi\right)
=\frac{m^2(r_+^2+a^2+n^2)(1-\Xi^2)}{\Xi K^4}\,,
\end{align}
which is nonzero for a generic rotating AdS solution.  The squared-mass
formula therefore requires $M=m/(\Xi K)$ in place of $\mathcal M$.
This check applies to the chosen mass formula and does not exclude a
different thermodynamic construction based on another energy.

\subsection{First law on the physical solution family}

The independent physical variations may be taken as
$(dr_+,da,dn,dq,dg,dK)$ after eliminating $dm$ with
$d\Delta_r(r_+)=0$.  Explicitly,
\begin{align}
2r_+\,dm &=\Delta_r'(r_+)\,dr_+
+\frac{\partial\Delta_r}{\partial a}\,da +\frac{\partial\Delta_r}{\partial n}\,dn
+2q\,dq +\frac{\partial\Delta_r}{\partial g}\,dg \, ,
\label{app:deltadiff}
\end{align}
with all derivatives evaluated at $r=r_+$.  Pulling back the differential of
Eq.~(\ref{smf}) with Eq.~(\ref{app:deltadiff}) gives
\begin{align}
&dM -TdS-\Omega dJ-\omega{} dJ_n-\Phi dQ-\Psi dN -VdP
 \nn \\
&+\lambda_+d\mu_+ +\lambda_-d\mu_-=0\,.
\end{align}
The same substitution gives the Euler relation
\begin{align}
M-2TS-2\Omega J-2\omega{}J_n-\Phi Q-\Psi N+2VP=0\,.
\end{align}
Together, these identities verify Eqs. (\ref{FL}) and (\ref{Smarr}) on the physical
family with the full parameter dependence, before any limiting family
is selected.


\begin{thebibliography}{99}


\bibitem{AnnMath53-472}
A.~H.~Taub,
Empty space-times admitting a three parameter group of motions,
\href{https://doi.org/10.2307/1969567}
{Annals Math. \textbf{53}, 472 (1951)}.

\bibitem{JMP4-915}
E.~Newman, L.~Tamburino, and T.~Unti,
Empty-space generalization of the Schwarzschild metric,
\href{https://doi.org/10.1063/1.1704018}
{J. Math. Phys. \textbf{4}, 915 (1963)}.

\bibitem{JMP4-924}
C.~W.~Misner,
The flatter regions of Newman, Unti and Tamburino's generalized Schwarzschild space,
\href{https://doi.org/10.1063/1.1704019}
{J. Math. Phys. \textbf{4}, 924 (1963)}.

\bibitem{CQG26-195011}
D.~Kastor, S.~Ray, and J.~Traschen,
Enthalpy and the mechanics of AdS black holes,
\href{https://doi.org/10.1088/0264-9381/26/19/195011}
{Class. Quant. Grav. \textbf{26}, 195011 (2009)}
[\href{https://arxiv.org/abs/0904.2765}{arXiv:0904.2765}].

\bibitem{CQG28-235017}
B.~P.~Dolan,
Pressure and volume in the first law of black hole thermodynamics,
\href{https://doi.org/10.1088/0264-9381/28/23/235017}
{Class. Quant. Grav. \textbf{28}, 235017 (2011)}
[\href{https://arxiv.org/abs/1106.6260}{arXiv:1106.6260}].

\bibitem{PRD84-024037}
M.~Cveti\v{c}, G.~W.~Gibbons, D.~Kubiz\v{n}\'ak, and C.~N.~Pope,
Black hole enthalpy and an entropy inequality for the thermodynamic volume,
\href{https://doi.org/10.1103/PhysRevD.84.024037}
{Phys. Rev. D \textbf{84}, 024037 (2011)}
[\href{https://arxiv.org/abs/1012.2888}{arXiv:1012.2888}].

\bibitem{JHEP0712033}
D.~Kubiz\v{n}\'ak and R.~B.~Mann,
$P$-$V$ criticality of charged AdS black holes,
\href{https://doi.org/10.1007/JHEP07(2012)033}
{J. High Energy Phys. \textbf{07} (2012) 033}
[\href{https://arxiv.org/abs/1205.0559}{arXiv:1205.0559}].

\bibitem{PRL25-1596}
D.~Christodoulou,
Reversible and irreversible transformations in black hole physics,
\href{https://doi.org/10.1103/PhysRevLett.25.1596}
{Phys. Rev. Lett. \textbf{25}, 1596 (1970)}.

\bibitem{PRD4-3552}
D.~Christodoulou and R.~Ruffini,
Reversible transformations of a charged black hole,
\href{https://doi.org/10.1103/PhysRevD.4.3552}
{Phys. Rev. D \textbf{4}, 3552 (1971)}.

\bibitem{PRD100-101501}
S.-Q.~Wu and D.~Wu,
Thermodynamical hairs of the four-dimensional Taub-Newman-Unti-Tamburino spacetimes,
\href{https://doi.org/10.1103/PhysRevD.100.101501}
{Phys. Rev. D \textbf{100}, 101501 (2019)}
[\href{https://arxiv.org/abs/1909.07776}{arXiv:1909.07776}].

\bibitem{PRD105-124013}
D.~Wu and S.-Q.~Wu,
Consistent mass formulas for the four-dimensional dyonic NUT-charged spacetimes,
\href{https://doi.org/10.1103/PhysRevD.105.124013}
{Phys. Rev. D \textbf{105}, 124013 (2022)}
[\href{https://arxiv.org/abs/2202.09251}{arXiv:2202.09251}].

\bibitem{PRD7-2333}
J.~D.~Bekenstein,
Black holes and entropy,
\href{https://doi.org/10.1103/PhysRevD.7.2333}
{Phys. Rev. D \textbf{7}, 2333 (1973)}.

\bibitem{PRD13-191}
S.~W.~Hawking,
Black holes and thermodynamics,
\href{https://doi.org/10.1103/PhysRevD.13.191}
{Phys. Rev. D \textbf{13}, 191 (1976)}.

\bibitem{2607.23644}
D.~Wu, R.~B.~Mann, and S.-Q.~Wu,
Two-horizon sector thermodynamics as a diagnostic for the bi-hair organization of NUT charge,
\href{https://arxiv.org/abs/2607.23644}{arXiv:2607.23644}.

\bibitem{PRD59-044033}
S.~W.~Hawking, C.~J.~Hunter, and D.~N.~Page,
Nut charge, anti-de Sitter space and entropy,
\href{https://doi.org/10.1103/PhysRevD.59.044033}
{Phys. Rev. D \textbf{59}, 044033 (1999)}
[\href{https://arxiv.org/abs/hep-th/9809035}{hep-th/9809035}].

\bibitem{PRD61-084013}
R.~B.~Mann,
Entropy of rotating Misner string space-times,
\href{https://doi.org/10.1103/PhysRevD.61.084013}
{Phys. Rev. D \textbf{61}, 084013 (2000)}
[\href{https://arxiv.org/abs/hep-th/9904148}{hep-th/9904148}].

\bibitem{hep-th/0305119}
T.~Ghosh,
Thermodynamics of a class of Kerr-Bolt-AdS space-time,
[\href{https://arxiv.org/abs/hep-th/0305119}{hep-th/0305119}].

\bibitem{1406.1257}
S.~MacDonald,
Thermodynamic volume of Kerr-bolt-AdS spacetime,
[\href{https://arxiv.org/abs/1406.1257}{arXiv:1406.1257}].

\bibitem{PRD100-064055}
R.~A.~Hennigar, D.~Kubiz\v{n}\'{a}k, and R.~B.~Mann,
Thermodynamics of Lorentzian Taub-NUT spacetimes,
\href{https://doi.org/10.1103/PhysRevD.100.064055}
{Phys. Rev. D \textbf{100}, 064055 (2019)}
[\href{https://arxiv.org/abs/1903.08668}{arXiv:1903.08668}].

\bibitem{JHEP0719119}
A.~B.~Bordo, F.~Gray, and D.~Kubiz\v{n}\'ak,
Thermodynamics and phase transitions of NUTty dyons,
\href{https://doi.org/10.1007/JHEP07(2019)119}
{J. High Energy Phys. \textbf{07} (2019) 119}
[\href{https://arxiv.org/abs/1904.00030}{arXiv:1904.00030}].

\bibitem{CQG36-194001}
A.~B.~Bordo, F.~Gray, R.~A.~Hennigar, and D.~Kubiz\v{n}\'{a}k,
Misner gravitational charges and variable string strengths,
\href{https://doi.org/10.1088/1361-6382/ab3d4d}
{Class. Quant. Grav. \textbf{36}, 194001 (2019)}
[\href{https://arxiv.org/abs/1905.03785}{arXiv:1905.03785}].

\bibitem{PLB798-134972}
A.~B.~Bordo, F.~Gray, R.~A.~Hennigar, and D.~Kubiz\v{n}\'{a}k,
The first law for rotating NUTs,
\href{https://doi.org/10.1016/j.physletb.2019.134972}
{Phys. Lett. B \textbf{798}, 134972 (2019)}
[\href{https://arxiv.org/abs/1905.06350}{arXiv:1905.06350}].

\bibitem{PRD100-104016}
Z.~Chen and J.~Jiang,
General Smarr relation and first law of a NUT dyonic black hole,
\href{https://doi.org/10.1103/PhysRevD.100.104016}
{Phys. Rev. D \textbf{100}, 104016 (2019)}
[\href{https://arxiv.org/abs/1910.10107}{arXiv:1910.10107}].

\bibitem{JHEP0520084}
A.~Ballon Bordo, F.~Gray, and D.~Kubiz\v{n}\'{a}k,
Thermodynamics of rotating NUTty dyons,
\href{https://doi.org/10.1007/JHEP05(2020)084}
{J. High Energy Phys. \textbf{05} (2020) 084}
[\href{https://arxiv.org/abs/2003.02268}{arXiv:2003.02268}].

\bibitem{JHEP1022044}
N.~H.~Rodr\'iguez and M.~J.~Rodriguez,
First law for Kerr Taub-NUT AdS black holes,
\href{https://doi.org/10.1007/JHEP10(2022)044}
{J. High Energy Phys. \textbf{10} (2022) 044}
[\href{https://arxiv.org/abs/2112.00780}{arXiv:2112.00780}].

\bibitem{IJMPD31-2250021}
R.~Durka,
The first law of black hole thermodynamics for Taub-NUT spacetime,
\href{https://doi.org/10.1142/S0218271822500213}
{Int. J. Mod. Phys. D \textbf{31}, 2250021 (2022)}
[\href{https://arxiv.org/abs/1908.04238}{arXiv:1908.04238}].

\bibitem{PLB802-135270}
G.~Cl\'ement and D.~Gal'tsov,
On the Smarr formulas for electrovac spacetimes with line singularities,
\href{https://doi.org/10.1016/j.physletb.2020.135270}
{Phys. Lett. B \textbf{802}, 135270 (2020)}
[\href{https://arxiv.org/abs/1908.10617}{arXiv:1908.10617}].

\bibitem{PRD106-024022}
M.~Godazgar and S.~Guisset,
Dual charges for AdS spacetimes and the first law of black hole mechanics,
\href{https://doi.org/10.1103/PhysRevD.106.024022}
{Phys. Rev. D \textbf{106}, 024022 (2022)}
[\href{https://arxiv.org/abs/2205.10043}{arXiv:2205.10043}].

\bibitem{JHEP0321039}
R.~B.~Mann, L.~A.~Pando Zayas, and M.~Park,
Complement to thermodynamics of dyonic Taub-NUT-AdS spacetime,
\href{https://doi.org/10.1007/JHEP03(2021)039}
{J. High Energy Phys. \textbf{03} (2021) 039}
[\href{https://arxiv.org/abs/2012.13506}{arXiv:2012.13506}].

\bibitem{PRD101-124011}
A.~Awad and S.~Eissa,
Topological dyonic Taub-Bolt/NUT-AdS: thermodynamics and first law,
\href{https://doi.org/10.1103/PhysRevD.101.124011}
{Phys. Rev. D \textbf{101}, 124011 (2020)}
[\href{https://arxiv.org/abs/2007.10489}{arXiv:2007.10489}].

\bibitem{PRD108-064022}
A.~Awad and E.~Elkhateeb,
Dyonic Taub-NUT-AdS: unconstrained thermodynamics and phase structure,
\href{https://doi.org/10.1103/PhysRevD.108.064022}
{Phys. Rev. D \textbf{108}, 064022 (2023)}
[\href{https://arxiv.org/abs/2304.06705}{arXiv:2304.06705}].

\bibitem{PRD109-084026}
M.~Tharwat, A.~AlBarqawy, A.~Awad, and E.~Elkhateeb,
Dyonic Taub-NUT-AdS spaces: phase structures of all horizon geometries,
\href{https://doi.org/10.1103/PhysRevD.109.084026}
{Phys. Rev. D \textbf{109}, 084026 (2024)}
[\href{https://arxiv.org/abs/2312.15811}{arXiv:2312.15811}].

\bibitem{EPJC85-1411}
A.~AlBarqawy, A.~Awad, E.~Elkhateeb, and M.~Tharwat,
Dyonic Taub-NUT-AdS black branes: thermodynamics and phase diagrams,
\href{https://doi.org/10.1140/epjc/s10052-025-15144-3}
{Eur. Phys. J. C \textbf{85}, 1411 (2025)}
[\href{https://arxiv.org/abs/2502.19511}{arXiv:2502.19511}].

\bibitem{PRD105-124034}
A.~Awad and S.~Eissa,
Lorentzian Taub-NUT spacetimes: Misner string charges and the first law,
\href{https://doi.org/10.1103/PhysRevD.105.124034}
{Phys. Rev. D \textbf{105}, 124034 (2022)}
[\href{https://arxiv.org/abs/2206.09124}{arXiv:2206.09124}].

\bibitem{JHEP1022174}
H.-S.~Liu, H.~L\"u, and L.~Ma,
Thermodynamics of Taub-NUT and Plebanski solutions,
\href{https://doi.org/10.1007/JHEP10(2022)174}
{J. High Energy Phys. \textbf{10} (2022) 174}
[\href{https://arxiv.org/abs/2208.05494}{arXiv:2208.05494}].

\bibitem{EPJC84-515}
J.-F.~Liu and H.-S.~Liu,
Thermodynamics of Taub-NUT-AdS spacetimes,
\href{https://doi.org/10.1140/epjc/s10052-024-12826-2}
{Eur. Phys. J. C \textbf{84}, 515 (2024)}
[\href{https://arxiv.org/abs/2309.01609}{arXiv:2309.01609}].

\bibitem{PLB846-138227}
D.~Wu and S.-Q.~Wu,
Revisiting mass formulas of the four-dimensional Reissner-Nordstr\"om-NUT-AdS solutions in a different metric form,
\href{https://doi.org/10.1016/j.physletb.2023.138227}
{Phys. Lett. B \textbf{846}, 138227 (2023)}
[\href{https://arxiv.org/abs/2210.17504}{arXiv:2210.17504}].

\bibitem{PRD108-064034}
D.~Wu and S.-Q.~Wu,
Consistent mass formulas for higher even-dimensional Taub-NUT spacetimes and their AdS counterparts,
\href{https://doi.org/10.1103/PhysRevD.108.064034}
{Phys. Rev. D \textbf{108}, 064034 (2023)}
[\href{https://arxiv.org/abs/2209.01757}{arXiv:2209.01757}].

\bibitem{PRD108-064035}
S.-Q.~Wu and D.~Wu,
Consistent mass formulas for higher even-dimensional Reissner-Nordstr\"om-NUT-AdS spacetimes,
\href{https://doi.org/10.1103/PhysRevD.108.064035}
{Phys. Rev. D \textbf{108}, 064035 (2023)}
[\href{https://arxiv.org/abs/2306.00062}{arXiv:2306.00062}].

\bibitem{CQG22-1503}
G.~W.~Gibbons, M.~J.~Perry, and C.~N.~Pope,
The First law of thermodynamics for Kerr-anti-de Sitter black holes,
\href{https://doi.org/10.1088/0264-9381/22/9/002}
{Class. Quant. Grav. \textbf{22}, 1503 (2005)}
[\href{https://arxiv.org/abs/hep-th/0408217}{hep-th/0408217}].

\bibitem{JHEP0517116}
M.~Appels, R.~Gregory, and D.~Kubiz\v{n}\'{a}k,
Black hole thermodynamics with conical defects,
\href{https://doi.org/10.1007/JHEP05(2017)116}
{J. High Energy Phys. \textbf{05} (2017) 116}
[\href{https://arxiv.org/abs/1702.00490}{arXiv:1702.00490}].

\bibitem{AP90-196}
J.~F.~Plebanski,
A class of solutions of Einstein-Maxwell equations,
\href{https://doi.org/10.1016/0003-4916(75)90145-1}
{Annals Phys. \textbf{90}, 196 (1975)}.

\bibitem{AP98-98}
J.~F.~Pleba\'nski and M.~Demia\'nski,
Rotating, charged, and uniformly accelerating mass in general relativity,
\href{https://doi.org/10.1016/0003-4916(76)90240-2}
{Annals Phys. \textbf{98}, 98 (1976)}.

\bibitem{IJMPD15-335}
J.~B.~Griffiths and J.~Podolsky,
A new look at the Plebanski-Demianski family of solutions,
\href{https://doi.org/10.1142/S0218271806007742}
{Int. J. Mod. Phys. D \textbf{15}, 335 (2006)}
[\href{https://arxiv.org/abs/gr-qc/0511091}{gr-qc/0511091}].

\bibitem{CQG22-3467}
J.~B.~Griffiths and J.~Podolsky,
Accelerating and rotating black holes,
\href{https://doi.org/10.1088/0264-9381/22/17/008}
{Class. Quant. Grav. \textbf{22}, 3467 (2005)}
[\href{https://arxiv.org/abs/gr-qc/0507021}{gr-qc/0507021}].

\bibitem{PRD73-104036}
W.~Chen, H.~L\"u, and C.~N.~Pope,
Mass of rotating black holes in gauged supergravities,
\href{https://doi.org/10.1103/PhysRevD.73.104036}
{Phys. Rev. D \textbf{73}, 104036 (2006)}
[\href{https://arxiv.org/abs/hep-th/0510081}{hep-th/0510081}].

\bibitem{PRD110-104021}
C.~Corral and R.~Olea,
Electric-magnetic duality of dyonic Kerr-Newman-NUT-AdS spacetimes,
\href{https://doi.org/10.1103/PhysRevD.110.104021}
{Phys. Rev. D \textbf{110}, 104021 (2024)}
[\href{https://arxiv.org/abs/2408.03901}{arXiv:2408.03901}].

\bibitem{CQG17-399}
M.~M.~Caldarelli, G.~Cognola, and D.~Klemm,
Thermodynamics of Kerr-Newman-AdS black holes and conformal field theories,
\href{https://doi.org/10.1088/0264-9381/17/2/310}
{Class. Quant. Grav. \textbf{17}, 399 (2000)}
[\href{https://arxiv.org/abs/hep-th/9908022}{hep-th/9908022}].

\bibitem{CMP31-161}
J.~M.~Bardeen, B.~Carter, and S.~W.~Hawking,
The Four laws of black hole mechanics,
\href{https://doi.org/10.1007/BF01645742}
{Commun. Math. Phys. \textbf{31}, 161 (1973)}.

\bibitem{PRL30-71}
L.~Smarr,
Mass formula for Kerr black holes,
\href{https://doi.org/10.1103/PhysRevLett.30.71}
{Phys. Rev. Lett. \textbf{30}, 71 (1973)},
\href{https://doi.org/10.1103/PhysRevLett.30.521}
{Erratum: Phys. Rev. Lett. \textbf{30}, 521 (1973)}.

\bibitem{CQG31-235003}
C.~V.~Johnson,
Thermodynamic volumes for AdS-Taub-NUT and AdS-Taub-Bolt,
\href{https://doi.org/10.1088/0264-9381/31/23/235003}
{Class. Quant. Grav. \textbf{31}, 235003 (2014)}
[\href{https://arxiv.org/abs/1405.5941}{arXiv:1405.5941}].

\bibitem{EPJC83-365}
D.~Wu,
Classifying topology of consistent thermodynamics of the four-dimensional neutral Lorentzian NUT-charged spacetimes,
\href{https://doi.org/10.1140/epjc/s10052-023-11561-4}
{Eur. Phys. J. C \textbf{83}, 365 (2023)}
[\href{https://arxiv.org/abs/2302.01100}{arXiv:2302.01100}].

\bibitem{EPJC83-589}
D.~Wu,
Consistent thermodynamics and topological classes for the four-dimensional Lorentzian charged Taub-NUT spacetimes,
\href{https://doi.org/10.1140/epjc/s10052-023-11782-7}
{Eur. Phys. J. C \textbf{83}, 589 (2023)}
[\href{https://arxiv.org/abs/2306.02324}{arXiv:2306.02324}].

\bibitem{PRL129-191101}
S.-W.~Wei, Y.-X.~Liu, and R.~B.~Mann,
Black Hole Solutions as Topological Thermodynamic Defects,
\href{https://doi.org/10.1103/PhysRevLett.129.191101}
{Phys. Rev. Lett. \textbf{129}, 191101 (2022)}
[\href{https://arxiv.org/abs/2208.01932}{arXiv:2208.01932}].

\bibitem{PRD107-024024}
D.~Wu,
Topological classes of rotating black holes,
\href{https://doi.org/10.1103/PhysRevD.107.024024}
{Phys. Rev. D \textbf{107}, 024024 (2023)}
[\href{https://arxiv.org/abs/2211.15151}{arXiv:2211.15151}].

\bibitem{PRD107-084002}
D.~Wu and S.-Q.~Wu,
Topological classes of thermodynamics of rotating AdS black holes,
\href{https://doi.org/10.1103/PhysRevD.107.084002}
{Phys. Rev. D \textbf{107}, 084002 (2023)}
[\href{https://arxiv.org/abs/2301.03002}{arXiv:2301.03002}].

\bibitem{PRD108-084041}
D.~Wu,
Topological classes of thermodynamics of the four-dimensional static accelerating black holes,
\href{https://doi.org/10.1103/PhysRevD.108.084041}
{Phys. Rev. D \textbf{108}, 084041 (2023)}
[\href{https://arxiv.org/abs/2307.02030}{arXiv:2307.02030}].

\bibitem{JHEP0624213}
D.~Wu, S.-Y.~Gu, X.-D.~Zhu, Q.-Q.~Jiang, and S.-Z.~Yang,
Topological classes of thermodynamics of the static multi-charge AdS black holes in gauged supergravities: novel temperature-dependent thermodynamic topological phase transition,
\href{https://doi.org/10.1007/JHEP06(2024)213}
{J. High Energy Phys. \textbf{06} (2024) 213}
[\href{https://arxiv.org/abs/2402.00106}{arXiv:2402.00106}].

\bibitem{PLB856-138919}
X.-D.~Zhu, D.~Wu, and D.~Wen,
Topological classes of thermodynamics of the rotating charged AdS black holes in gauged supergravities,
\href{https://doi.org/10.1016/j.physletb.2024.138919}
{Phys. Lett. B \textbf{856}, 138919 (2024)}
[\href{https://arxiv.org/abs/2402.15531}{arXiv:2402.15531}].

\bibitem{PDU46-101617}
H.~Chen, D.~Wu, M.-Y.~Zhang, H.~Hassanabadi, and Z.-W.~Long,
Thermodynamic topology of phantom AdS black holes in massive gravity,
\href{https://doi.org/10.1016/j.dark.2024.101617}
{Phys. Dark Univ. \textbf{46}, 101617 (2024)}
[\href{https://arxiv.org/abs/2404.082431}{arXiv:2404.08243}].

\bibitem{CQG42-125007}
W.~Liu, L.~Zhang, D.~Wu, and J.~Wang,
Thermodynamic topological classes of the rotating, accelerating black holes,
\href{https://doi.org/10.1088/1361-6382/ade35b}
{Class. Quant. Grav. \textbf{42}, 125007 (2025)}
[\href{https://arxiv.org/abs/2409.11666}{arXiv:2409.11666}].

\bibitem{PLB860-139163}
X.-D.~Zhu, W.~Liu, and D.~Wu,
Universal thermodynamic topological classes of rotating black holes,
\href{https://doi.org/10.1016/j.physletb.2024.139163}
{Phys. Lett. B \textbf{860}, 139163 (2025)}
[\href{https://arxiv.org/abs/2409.12747}{arXiv:2409.12747}].

\bibitem{PRD110-L081501}
S.-W.~Wei, Y.-X.~Liu, and R.~B.~Mann,
Universal topological classifications of black hole thermodynamics,
\href{https://doi.org/10.1103/PhysRevD.110.L081501}
{Phys. Rev. D \textbf{110}, L081501 (2024)}
[\href{https://arxiv.org/abs/2409.09333}{arXiv:2409.09333}].

\bibitem{PRD111-L061501}
D.~Wu, W.~Liu, S.-Q.~Wu, and R.~B.~Mann,
Novel topological classes in black hole thermodynamics,
\href{https://doi.org/10.1103/PhysRevD.111.L061501}
{Phys. Rev. D \textbf{111}, L061501 (2025)}
[\href{https://arxiv.org/abs/2411.10102}{arXiv:2411.10102}].

\bibitem{PLB865-139482}
Y.~Chen, X.-D.~Zhu, and D.~Wu,
Universal thermodynamic topological classes of three-dimensional BTZ black holes,
\href{https://doi.org/10.1016/j.physletb.2025.139482}
{Phys. Lett. B \textbf{865}, 139482 (2025)}
[\href{https://arxiv.org/abs/2504.10858}{arXiv:2504.10858}].

\bibitem{EPJC85-828}
H.~Chen, D.~Wu, M.-Y.~Zhang, S.~Zare, H.~Hassanabadi, B.~C.~L\"utf\"uo\v{g}lu, and Z.-W.~Long,
Universal thermodynamic topological classes of static black holes in Conformal Killing Gravity,
\href{https://doi.org/10.1140/epjc/s10052-025-14581-4}
{Eur. Phys. J. C \textbf{85}, 828 (2025)}
[\href{https://arxiv.org/abs/2506.03695}{arXiv:2506.03695}].

\bibitem{PRD112-124024}
W.~Ai and D.~Wu,
$\widetilde{W}^{1+}$ subclass: Extending the topological classification of black hole thermodynamics,
\href{https://doi.org/10.1103/dy7y-j24r}
{Phys. Rev. D \textbf{112}, 124024 (2025)}
[\href{https://arxiv.org/abs/2509.03308}{arXiv:2509.03308}].

\bibitem{EPJC86-187}
D.~Wu and S.-Q.~Wu,
Thermodynamics and topological classifications of static non-extremal four-charge AdS black hole in the five-dimensional $\mathcal{N} = 2$, $STU-W^2U$ gauged supergravity,
\href{https://doi.org/10.1140/epjc/s10052-026-15379-8}
{Eur. Phys. J. C \textbf{86}, 187 (2026)}
[\href{https://arxiv.org/abs/2510.20164}{arXiv:2510.20164}].

\bibitem{2602.05231}
M.~Tian, Y.~Chen, and D.~Wu,
Dimensional structure of thermodynamic topology in ultraspinning Kerr-AdS black holes,
\href{https://doi.org/10.1016/j.physletb.2026.140658}
{Phys. Letts. B \textbf{879}, 140658 (2026)}
[\href{https://arxiv.org/abs/2602.05231}{arXiv:2602.05231}].

\bibitem{2605.00037}
S.-W.~Wei and Y.-X.~Liu,
Topology of black hole thermodynamics: A brief review,
\href{https://doi.org/10.1007/s11433-025-2923-3}
{Sci. China Phys. Mech. Astron. \textbf{69}, 260401 (2026)}
[\href{https://arxiv.org/abs/2605.00037}{arXiv:2605.00037}].

\end{thebibliography}
\end{document}